\documentclass[12pt,preprint]{aastex}

\newcommand{\kms}{${\rm km~s}^{-1}$}
\usepackage{lscape}
\slugcomment{Submitted to AJ}
\shorttitle{No excess of RR Lyrae stars in the CMa Overdensity}
\shortauthors{Mateu et al.}

\begin{document}

\title{No Excess of RR Lyrae Stars in the Canis Major Overdensity}

\author{Cecilia Mateu\altaffilmark{1} and A. Katherina Vivas}
\affil{Centro de Investigaciones de Astronom{\'\i}a (CIDA), Apartado Postal
264, M\'erida 5101-A, Venezuela}
\email{cmateu@cida.ve, vivas@cida.ve}

\author{Robert Zinn and Lissa R. Miller}
\affil{Department of Astronomy, Yale University, PO Box 208101, New Haven,
CT 06520-8101, USA}
\email{robert.zinn@yale.edu, miller@astro.yale.edu}

\and

\author{Carlos Abad}
\affil{Centro de Investigaciones de Astronom{\'\i}a (CIDA), Apartado Postal
264, M\'erida 5101-A, Venezuela}
\email{abad@cida.ve}

\altaffiltext{1}{Also at Universidad Central de Venezuela, Caracas}

\begin{abstract}

Our multi-epoch survey of $\sim 20$  sq. deg. of the Canis Major overdensity has detected only 10 RR Lyrae stars (RRLS).  We show that this number is consistent with the number expected from the Galactic halo and thick disk populations alone, leaving no excess that can be attributed to the dwarf spheroidal (dSph) galaxy that some authors have proposed as the origin of the CMa overdensity.  If this galaxy resembles the dSph satellites of the Milky Way and of M31 and has the putative $M_{V} \sim -14.5$, our survey should have detected several tens of RRLS.  Even if $M_{V} \lesssim-12$, the expected excess is  $\gtrsim10$, which is not observed.  Either the old stellar population of this galaxy has unique properties or, as others have argued before, the CMa overdensity is produced by the thin and thick disk and spiral arm populations of the Milky Way and not by a collision with a dSph satellite galaxy.   

\end{abstract}


\section{INTRODUCTION}\label{s:intro}

The Canis Major overdensity was discovered by \citet{mar04a}
as an excess in the density of 2MASS M-giant stars in the direction $(l,b)=(240\degr,-8\degr)$ 
compared to similar fields in the northern galactic hemisphere. It is estimated to span $\sim100$ sq.deg. on the sky
and to be concentrated at a distance of $\sim7.2$ kpc from the Sun \citep{mar04b,bel06}.
The large numbers of M-giant and red clump (RC) stars in the CMa overdensity suggest
that its main population is of intermediate age $\sim7$ Gyr \citep{bel04}. 
A "Blue Plume" of stars in the color-magnitude diagram (CMD) has been interpreted by \citet{bel04}
and \citet{m-d05} 
as the main-sequence of a younger population (age $\lesssim1$Gyr) 

The CMa overdensity was interpreted by \citet{mar04a}
as the remnant core of a disrupting dwarf galaxy.
However, its origin remains a topic of considerable debate.
The dwarf galaxy hypothesis has been contested by several authors 
who, using different tracers and techniques, have proposed that the overdensity 
is produced by the stellar populations of the warped Galactic thin disk \citep{mom04}, 
the thick disk \citep{car08}, and the outer Norma-Cygnus spiral arm \citep{moi06, car05}.  
Counter arguments have been made by \citet{mar04b,mar05} and \citet{bel06},
in support of the dwarf galaxy nature of CMa, claiming that the
Galactic warp is not sufficient to account for the overdensity.
The proper motions of a sample of blue plume stars have been measured by \citet{din05}, 
who concluded that these stars do not have the space motions expected of galactic stars, if they have 
the same radial velocity as the sample of M-giants that \citet{mar04b} attributed to the CMa dwarf galaxy.

The lack of consensus on this issue is probably a consequence
of our limited knowledge of the complex Galactic structure in
this direction and also the large and highly variable interstellar extinction. 
CMa lies in the third Galactic quadrant where the warp of the thin disk  
has its maximum amplitude \citep[at $l\sim270^0$,][]{mom04,l-c06} and where the Galactic 
thick disk might be warped as well \citep{car08}. Also, in this quadrant the spiral arms 
have not been traced as well as in other directions \citep{car05,mcc05}. 

In Table \ref{t:cmastudies}, we summarize the previous studies of the stellar populations in the
CMa overdensity. Most are photometric and use techniques such as main sequence (MS)
fitting and modelling of either Hess diagrams (HD) or CMDs to obtain the age,
metallicity and distance for the stellar population of the CMa overdensity. \citet{car05,moi06,car08} and 
\citet{pow08} have made 
use of multiband photometry and two color
diagrams (TCD) to reduce the effects of reddening.
The top part of the table shows the studies of the young population in the
region, while the bottom part indicates the intermediate age ones. 
As one can see from the table, no old population ($\gtrsim10$Gyr)
has been identified.

In summary, there are two conflicting views of the nature of the CMa overdensity:
\begin{itemize}
\item The CMa overdensity is a dSph galaxy undergoing tidal disruption.  The stellar population of this galaxy is dominated by an intermediate-age population, and there is a minority population of much younger stars \citep{mar04a,mar04b,bel06,m-d05}. 

\item The overdensity in the intermediate-age population is produced by a projection of the warped thin and thick disks \citep{mom04,l-c06,car08} into the line of sight. The young population 
corresponds to the outer part of the Norma-Cygnus spiral arm which, 
being part of the thin disk, also follows the warp \citep{car05,moi06,con07}. 
\end{itemize}

The goal of our work was to conduct a large scale survey of
RR Lyrae stars (RRLS) in order to trace the oldest stellar population ($\gtrsim10$Gyr)\footnote{Earlier progress reports of our survey can be found in \citet{mateu06,mateu07}}.  
Old stellar populations containing RRLS have been identified in every dSph satellite of the Milky Way (MW) and of M31 that has been properly surveyed \citep[see e.g.][]{viv06}.  If the CMa galaxy is similar to these systems in its stellar mix and has the proposed luminosity, a large excess ($>100$) of RRLS is expected (see \S 4).  Because RRLS are excellent standard candles, the distances to individual RRLS can be obtained with high precision ($\lesssim10\%$).  If there is a dSph galaxy at a distance of $\sim7.2$ kpc \citep{mar04a}, an overdensity of RRLS is expected at magnitudes $V\sim15$. The detection of an overdensity of RRLS that spans a narrow range of magnitudes would be conclusive proof that a dSph galaxy produces the CMa overdensity.  On the other hand, the absence of an excess of RRLS would weaken the dwarf galaxy hypothesis, leaving more likely a Galactic origin for the CMa overdensity. 

\section{PHOTOMETRIC OBSERVATIONS}\label{s:obs}

The large field of view ($2\fdg 2\times 2\fdg 3$) provided by the QUEST-I camera \citep{bal02},
installed on the 1m Jurgen Stock Telescope at the Venezuelan National Observatory of Llano
del Hato, made this an ideal instrument to conduct a large-scale, multi-epoch survey
of an extended structure such as the CMa overdensity. The observations were made in
the V band and were obtained in two runs during 2004 and 2005, centered 
at $l\backsimeq240\degr, b\backsimeq-9.5\degr$ and spanning $\sim 30$ sq. degrees
with an effective coverage of $\sim 20$sq. deg. (which takes into account the area
missed because of the gaps between the CCDs in the mosaic camera and the saturation of very bright stars). On average, we obtained $\sim19$ epochs
per field in $24$ nights, with a time spacing between consecutive observations ranging
from a few hours up to several weeks. Infrared J, H and K magnitudes from the 2MASS catalogue were
also obtained for each star.

\begin{figure}
\plotone{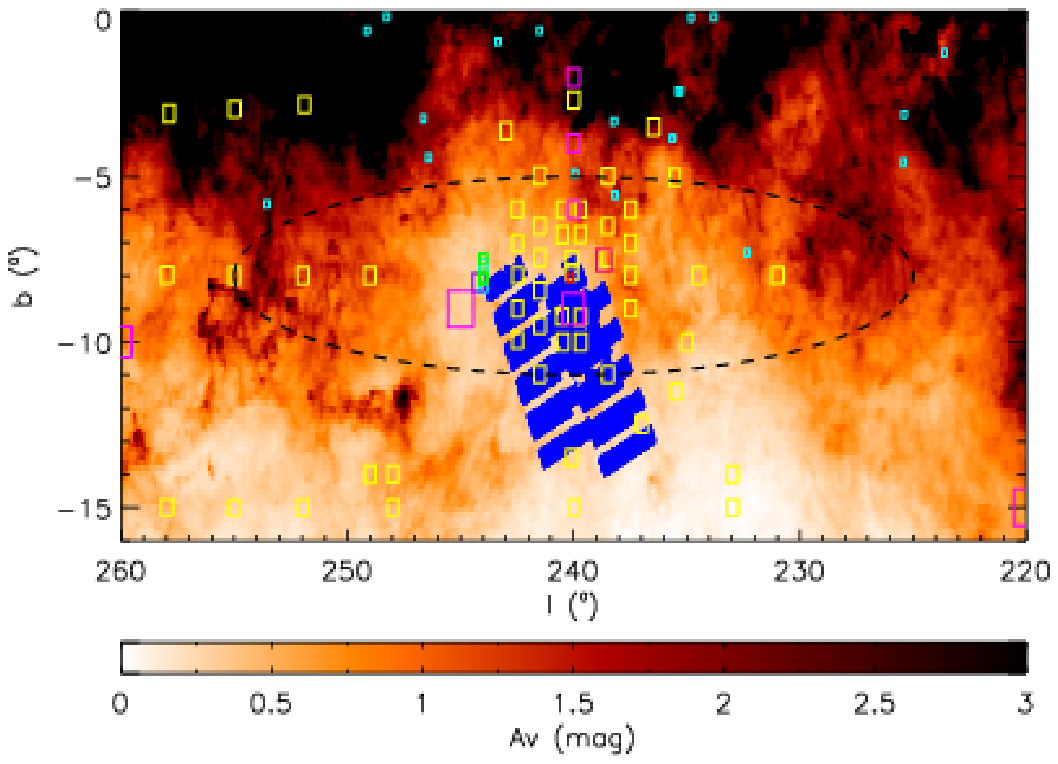}
\caption{Schematic plot of the areas in CMa studied by different authors.  
The blue area shows the QUEST RRLS survey footprint. 
The colored rectangles represent the areas studied by Bellazzini et al 2006 (purple), Martinez-Delgado
et al. 2005 (red), Carraro et al. 2005 and Moitinho et al. 2006 (cyan), 
Butler et al. 2007 and de Jong et al. 2007 (yellow), Conn et al. 2007 (magenta), Powell et al. 2008 
(dark pink) and Carraro et al. 2008 (green).
The axes of the dashed-lined ellipse are equal to the full-width at half maxima derived for CMa
using an aspect ratio of $5:1$ and the full-width at half maximum in the
$b$ direction measured by \citet{but07}. As noted by \citet{but07}, the aspect ratio
of CMa could be larger than $5:1$, in which case the ellipse would be more elongated in longitude.
The colorscale represents the $A_V$ values obtained from interpolation of the \citet{sch98} dust maps.}
\label{f:regionmaps}
\end{figure}

The area covered by our survey and previous studies is shown in
Fig. \ref{f:regionmaps}, where a schematic diagram is overlaid on an extinction map derived from the \citet{sch98} dust maps. The CMa feature spans $\sim100$ sq. deg. and has a smoothly varying density across the plane of the sky, which according to
\citet{mar04a} can be fitted with Gaussian profiles. 
The axes of the ellipse shown in Fig.\ref{f:regionmaps}
correspond to the full-width at half maxima (FWHM) derived for CMa
using the aspect ratio of $\sim5:1$ and the FWHM
in the latitude direction measured by \citet{but07}.
Assuming a Gaussian profile with these FWHM we computed
that $\sim3\%$ of the stars in CMa overdensity
should be present in our survey's footprint.  The percentage could be smaller than this because
\citet{but07} considered $5:1$ to be a lower limit of the aspect ratio.
From Fig. 6 of \citet{mar04a},
we estimate that the aspect ratio is probably not much larger
than $\sim6:1$, and consequently, that our survey's coverage is $\gtrsim2\%$.  In the following discussion, we adopt $2\%$ to be conservative.
As shown in Fig. \ref{f:regionmaps}, our survey includes the same central regions where others (e.g., Martinez-Delgado et al. 2005; Bellazzini et al.2006; Butler et al. 2007) have detected large overdensities of stars.
One can also see that the extinction is highly variable in the vicinity of CMa and that its
main body is located in a low extinction window, as noted by \citet{r-p06}.

\subsection{Image Reduction, Photometry and Astrometry}\label{s:imred}

The image frames were reduced using standard \emph{IRAF}\footnote{IRAF is distributed
by the National Optical Astronomy Observatories, which are operated by the Asociation of Universities for
Research in Astronomy Inc., under cooperative agreement with the National Science Foundation}
routines for bias subtraction and flat-fielding.  Due to the low galactic latitude 
of the surveyed region, the fields are highly crowded,
thus requiring the computation of magnitudes through PSF photometry.
The PSF photometry was obtained using \emph{DAOPHOT} tasks
with a model PSF spatially variable to second order, constructed
from $\sim130-150$ stars per image. 

The astrometric solution was computed for each image using the
program CM1 \citep{sto81}, which calculates the astrometric
matrix based on the coordinates of $\sim150$ catalogue stars per image, identified
from the USNOB1.0 catalogue \citep{mon03}. The precision of the
obtained astrometric solutions was $0\farcs18$, which was calculated by matching
the coordinates of all objects with the 2MASS point source catalogue,
and corresponds to the mean difference between
the CM1 and the 2MASS coordinates for each object. This precision is 
sufficient for identifying each star in the different observations.

\subsection{Photometric Calibration and Extinction Correction}\label{s:photcal}

The photometric calibration was made in two steps, the normalization of the instrumental
magnitudes with respect to a reference night, and the calibration of the reference 
night photometry using secondary photometric standards.

\begin{figure}
\plotone{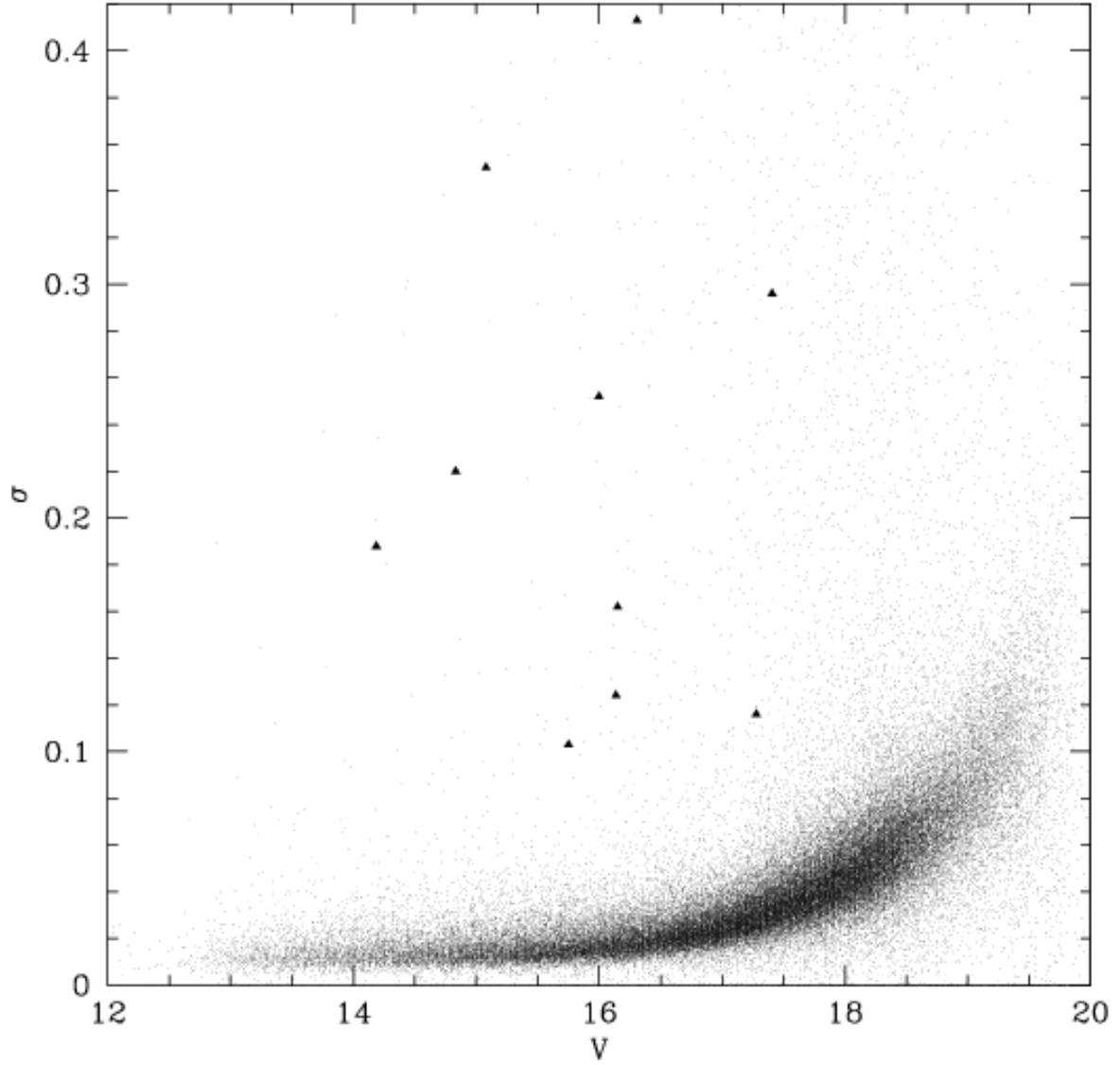}
\caption{Typical standard deviation of PSF magnitudes versus mean magnitude, based in $\sim20$ observations of $\sim80000$ stars. Solid triangles represent the positions of the RRLS identified in this work.}\label{f:e_vs_mag}
\end{figure}

In the first step, the magnitude zero-point of each night with respect
to the reference night was obtained by calculating a 3-sigma clipped mean of the magnitude 
differences of the stars that were detected on both nights. The use of $\sim 2000$ stars in the calculations made these zero-points statistically robust. 
The obtained zero-point was then added to the instrumental magnitudes of all the objects observed
on that particular night. At the completion of this process, mean photometric errors were evaluated as a function of magnitude by calculating the standard deviation of the multi-epoch observations 
of each object. A typical plot of standard deviation $\sigma$ of PSF magnitudes as a function of the mean magnitude is shown in Fig. \ref{f:e_vs_mag}.

For the second step, observations of the secondary standard fields were made
with the 1.3m telescope of the SMARTS consortium at the 
Cerro Tololo Interamerican Observatory (CTIO), Chile, with the $2048 \times 2048$ pixel
Fairchild 447 CCD camera, which has a field of view of $6' \times 6'$. The observations
were made in the R and V filters on 2004 January 23, under photometric conditions. 
A field containing Landolt standards was also observed during the night at different airmasses. 
The photometric solution for the night was
applied to the objects detected in the secondary standard fields.  The $316$ stars that were selected as secondary
standards have photometry in both filters and photometric errors $\leq 0.015$
magnitudes.

The secondary standards fields were chosen to be centered on open clusters lying in the 
region, to ensure having blue stars among the secondary standards. Consequently, the color distribution 
of the $316$
secondary standards was quite broad, $0.1 < {\rm V-R} < 0.8$.
Each row of CCDs in the QUEST camera was calibrated independently.
The mean \emph{rms} of the calculated solutions is $0.03$ magnitudes. 

The limiting magnitude obtained is $V_{lim}=19.5$, defined as the typical
magnitude of objects with $\sigma=0.1$ mag. The completeness magnitude obtained is $V_{com}=18.5$,
which corresponds to the mode of the magnitude distribution of all objects. Finally, the saturation 
magnitude of the survey is $V_{sat}\sim13.5$. The heliocentric distance range explored
by our RRLS survey is therefore $3$ kpc $\leq D_\odot\leq 39$ kpc (for the complete sample and assuming $M_V=+0.55$), 
which clearly includes the distance range of the CMa overdensity 
($5$ kpc $\lesssim D_\odot \lesssim10$ kpc, see Table \ref{t:cmastudies}).

Finally, the extinction correction was performed on a star by star basis, since extinction is highly variable
in the surveyed region (as shown in Fig. \ref{f:regionmaps}) due to its proximity to the Galactic plane. 
This was done by subtracting the corresponding $A_V$ at the position of each star, computed
from a linear interpolation of the \citet{sch98} dust maps, and including the asymptotic correction
proposed by \citet{bon00}. 

\section{RR LYRAE STARS}\label{s:rrls}

\subsection{RR Lyrae search}\label{s:rrlsearch}

The first step in the RRLS search was the identification of stars showing photometric 
variability in the V magnitude, using the procedure explained in \citet{viv04}. 
This consisted of computing, via a $\chi^2$ test,
the probability of the V magnitude dispersion being solely due to the photometric errors. From the catalogue of $286,837$ stars, $2775$ stars had $P_{\chi^2} \leq 0.001$
which indicates that they have a probability of $99.9\%$ of being variable.

The RRLS search was further restricted to the $651$ variable candidates fulfilling the
following criteria: (i) intrinsic color indices $(V-J)_0, (V-H)_0, (V-K)_0$ consistent with
those of spectral types earlier than G0. 
(ii) observed amplitude $\Delta V\geq 0.20$, and (iii)
number of observations in the V filter, $n_{obs}^V \geq 10$.

The RRLS search followed the procedure devised by \citet{lay98} for fitting the light curves of RRLS.
This consisted of fitting light-curve templates while varying four parameters: period,
amplitude, phase offset and magnitude at maximum light. 
For the fits of RR\emph{ab} variables, we used the six templates made by \citet{lay98}, which cover the range of light-curve shapes of this type of star. For RR\emph{c} stars, we used
either a cosine curve or a template made by averaging and smoothing the light curves of
10 well-observed RR{\it c} stars from the QUEST RR Lyrae catalogue \citep{viv04}.

The ranges for the trial parameters were explored according to the known distribution of periods 
and amplitudes for RRLS of each type \citep{viv04}, using a step of $10^{-5}$ d for the period in the range $0.4-1.2$ d
for RR$ab$ and $0.2-0.5$ d for RR$c$;
a step of $0.01$ mag for the amplitude in the ranges $0.2-1.4$ mag for RR$ab$ and $0.18-0.7$ for RR$c$;
and steps of $0.01$ mag and $0.005$ for the V magnitude at maximum light and 
the phase offset respectively. These values were then taken as the uncertainties
of the corresponding parameters. For each set of trial parameters, 
the $\chi^2$ of the fit was computed and finally, the four best fits for each RRLS type were 
visually inspected for every star. The stars for which an adequate template fit was found were
selected as initial RRLS candidates. Additionally, the epoch at maximum light and the flux-averaged V 
magnitude $\langle V \rangle$ were calculated using the best fitting light-curve parameters.

\subsection{Photometric Follow-up}\label{s:photfollowup}

The observations with the Stock telescope yielded for some RRLS candidates poorly sampled light curves. Consequently, we obtained additional observations with the $0.9$m, $1.0$m and $1.3$m telescopes of the SMARTS consortium at Cerro Tololo Interamerican Observatory  (CTIO), during Sept-Oct 2004 and Jan-Feb 2008;
and with the 1.0m Reflector at the National Observatory of Venezuela (NOV), during Jan 2008.
All observations were made in the V filter. This photometry 
was normalized to the reference night of the QUEST photometry using $\sim120-250$ field stars
per image for the CTIO observations and $\sim70-80$ for the NOV observations. 
On average, $5$ additional epochs were obtained for each candidate RRLS, 
yielding an average $24$ observations per star.  
Light curve parameters were recalculated with the additional observations,
including exploration of period aliases.
With these additional observations, only $16$ stars passed our tests for RRLS candidates, 
$10$ of type $c$ and $6$ of type ab. 

\subsection{Spectroscopic Follow-up}\label{s:specfollowup}

Spectroscopic follow-up of the 14 brighther ($V<16.5$, equivalent to $D_\odot\lesssim 12$ kpc)
RRLS was conducted in order to obtain
radial velocities and metallicities. These bright stars lie within the distance range
of the CMa overdensity.

Spectra were obtained with the Cassegrein spectrograph at the 1.5m SMARTS telescope
in Cerro Tololo Interamerican Observatory, Chile. The observations were obtained in service
mode in the months 2004 November-December, 2005 January, and 2008 February. The instrumental setup
was the same used by our group to observe a large number of halo RRLS from the QUEST survey.
We refer the reader to \citet{viv08} for details on the instrumental setup, data
reduction, and methods to estimate radial velocities and metallicities. Here we will
summarize only the more important issues. The spectra have a resolution of 4.3 {\AA } and a spectral
range from 3500 to 5300 \AA. Radial velocities were obtained by Fourier cross-correlation
(IRAF's {\sl fxcor} task) with several radial velocity standards of the same spectral type.
The radial velocity standards were chosen from the list of \citet{lay94} which also 
serve as pseudo-equivalent width standards for estimating the metallicities. Each RRLS was observed
twice at different phases during the pulsation cycle. To obtain the systemic velocity,
$V_\gamma$, of each
star, the radial velocity curve of the well-studied star X Arietis was fitted to the data 
\citep[see also ][]{viv05}.
The metallicities were estimated using the calibration by
\citet{lay94} of the pseudo-equivalent width of the Ca II K line
(W(K)), corrected for interstellar absorption, and the mean pseudo-equivalent
widths of the $\beta$, $\gamma$, and $\delta$ Balmer lines (W(H)).
The estimated error of the [Fe/H] measurements, which are on the \citet{zin84}
metallicity scale, is 0.15 dex for the RR$ab$ and
0.20 dex for the RR$c$. 

\subsection{Contaminants of the Sample}\label{s:contamination}

Our sample is expected to be contaminated by other types of variable stars.  In order of decreasing likelihood, these are contact eclipsing binaries (W UMa stars ) of A-F spectral type, pulsating blue straggler stars of the Delta Scuti and SX Phoenicis types, and Cepheids of type II and of the anomalous classification.  

In many cases the light curves of W UMa stars resemble a cosine curve \citep{ruc93}, 
making it difficult to distinguish them photometrically from RR$c$ stars,
since their period ranges also overlap. However, several orbital 
configurations do have light curves that can be distinguished, and
consequently, we also fitted a W UMa templates to some of the candidate RR$c$ stars.
This template was
constructed from the Fourier coefficientes of \citet{ruc93} ($i=80^0$,$q=0.4$,$f=0$), following
\citet{lay98}. We used only this W UMa template, which we consider representative of the type, 
since our goal was only to reduce the contamination that these stars introduce in the RR$c$ sample,
and not to survey their properties.  The W UMa template yielded a better fit than the RR$c$ template for $3$ stars, namely 25711, 10651 and 18056. The radial velocities of these stars are also consistent with the interpretation that these stars are not RRLS.
The radial velocities of stars 25711 and 10651 appear to be constant, while for star 18056 the radial velocity curve 
template for RRc stars does not fit the observations properly. Furthermore, the mean velocities of these 3 stars are $81$ \kms, $26$ \kms
and $77$ \kms, respectively, which lie close to the radial velocity peak expected for thin disk stars, which for the survey region is $V_\gamma\sim69$ {\kms} with $\sigma V_\gamma\sim22$ {\kms} (see \S \ref{s:radvel}). 
We conclude that stars 25711, 10651 and 18056 are most likely eclipsing 
binaries from the thin disk, and have classified them as W UMa stars. 

In the selection of RRLS candidates, we imposed color cuts to exclude the numerous red variables in the fields, but some non-RRLS may pass these criteria.  Our spectroscopy revealed that the RR$c$ candidates 6460, 5564, and 17237 have late F spectral types.  Since these types are too late for RR$c$ stars (Smith 1995), we have removed them from the sample of candidate RRLS. 

The Delta Scuti and SX Phoenicis variables, also known as dwarf cepheids,
have spectral types A0-F5 III-V, V amplitudes $<0.7$ mag and periods in the
ranges $0.01-0.2$ d and $0.04-0.08$ d respectively. 
Consequently, the period range of Delta Scuti stars overlaps with the short period
end of the RR$c$ period distribution \citep{smi95,kho98}. In the case of SX Phoenicis stars,
it is expected that some stars will have period aliases 
within the range expected for RR$c$ stars, which may escape detection by the low number 
of epochs and the time spacing of our observations. 
Although much less numerous than eclipsing binaries, Delta Scuti and SX Phoenicis stars
are expected in the Galactic disk and halo respectively, and hence are possible contaminants of the RR$c$ sample. 

Finally, a possible contaminant of the RR$ab$ sample are Anomalous Cepheids and short period type II Cepheids, also known as BL Herculis and W Virginis stars \citep{san06,kho98}. 
These stars have light curve shapes and period ranges that partially overlap with RRLS stars. 
However, these stars are very rare in comparison to the RRLS in the halo population, which is the only MW population in which they have been found.  Given that the halo population is itself a  small contributor to the stellar density in the CMa fields (see \S 5), the contamination from these stars should be small compared to the other types of stars.

\begin{figure}
\plotone{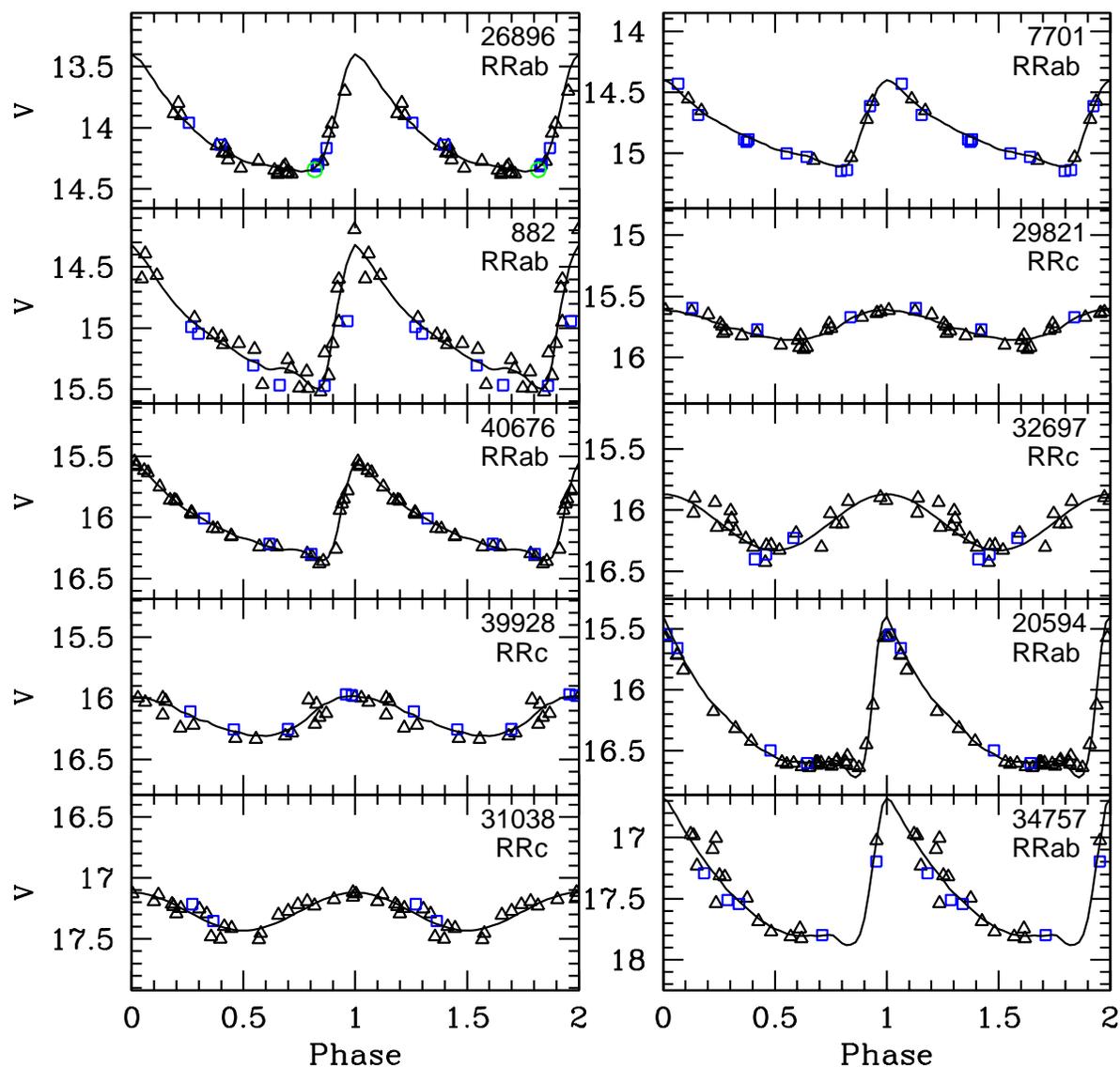}
\caption{The light curves of the identified RRLS. Observations made with the QUEST camera are shown as open triangles, and follow-up observations from the SMARTS telescopes and 
the NOV 1.0m Reflector  are shown with open squares and circles respectively. Solid lines show the best fitting template. Typical
error bars are smaller than the symbol size.}
\label{f:rrlcs}
\end{figure}

\begin{figure}
\includegraphics[bb=20 530 1900 800, clip, scale=0.85]{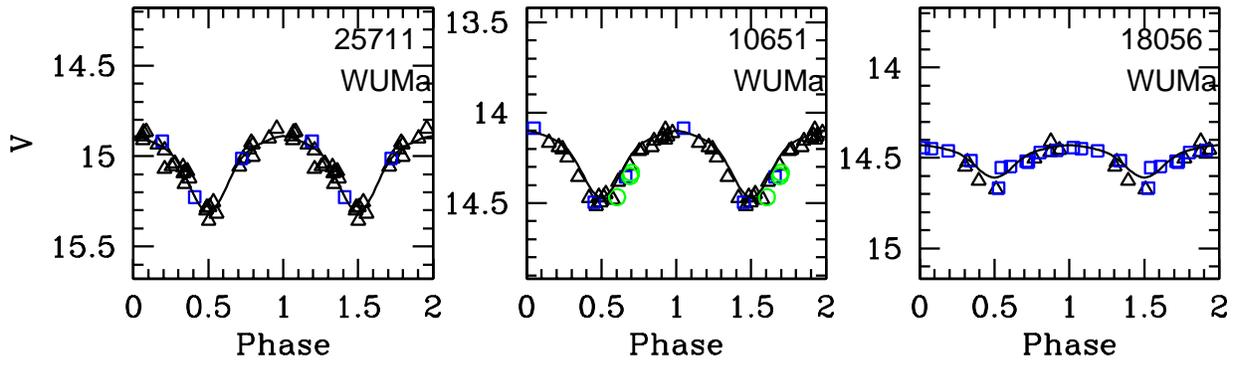}
\caption{The light curves of the identified W UMa eclipsing contact binaries. Symbols are as in Fig. \ref{f:rrlcs}.
Solid lines show the best fitting template.}
\label{f:wuma_lcs}
\end{figure}

After the removal of the likely contaminants, $10$ stars remained in our sample of RRLS candidates.
The light curves for the $10$ RRLS are shown in Figure \ref{f:rrlcs}, and those of the $3$ W UMa
stars are shown in Figure \ref{f:wuma_lcs}. The light curve and physical parameters obtained for the 
$10$ RRLS in Table \ref{t:rrpars}, 
including heliocentric distances of RRLS calculated based on the flux-averaged mean
magnitude $\langle V_0\rangle$ and assuming $M_V^{RR}=+0.55$ \citep{dem00,viv01}.
The uncertainty in the distance to the RRLS has been estimated in 7\%, following \citet{viv06}. 
The light curve parameters of the rejected RRLS candidates are summarized in Table \ref{t:rejpars}.
We must caution, however, that some contamination by non-RRLS may be still present, 
particularly for the sample of RR$c$ stars. 

\subsection{Completeness of the Survey}\label{s:completeness}

\begin{figure}
\plotone{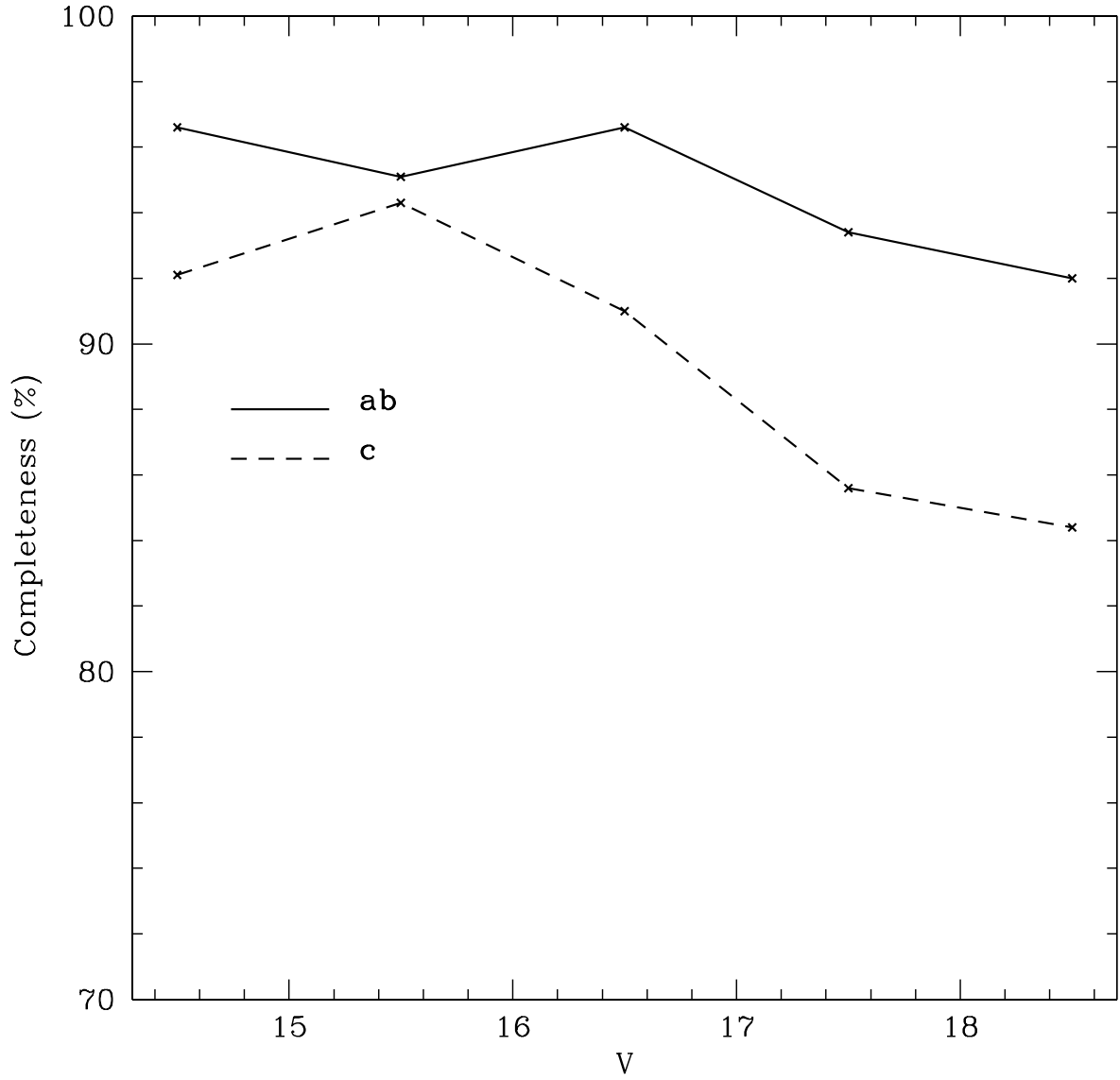}
\caption{Completeness of the survey as a function of V magnitude.}
\label{f:comp}
\end{figure}

The survey's completeness was evaluated as the fraction of synthetically generated light curves
for which the period was succesfully recovered.
A total of $3600$ synthetic RRLS light curves were generated for both RR$ab$ and RR$c$ types,
with $10$, $15$ and $20$ data points and a time sampling characteristic of the survey. Typical random
photometric errors were added as a function of magnitude (Fig. \ref{f:e_vs_mag}).
The synthetic light curves were used as input for the RRLS
identification algorithm described in \S \ref{s:rrlsearch}. If for a synthetic light curve,
the difference between the real and the best fitting period was better than $1\%$, it was considered as recovered.
The percentage of recovered light curves as a function of magnitude
for types RR$ab$ and RR$c$ are shown in Fig. \ref{f:comp}.  For this plot, we averaged the results obtained for light curves with different numbers of data points. One can see from this figure that the average completeness of the survey is $\sim90\%$.

RRLS 26896 has been previously identified as an RRLS by the All Sky
Automated Survey \citep[ASAS-3,][]{poj02}.
The period derived for this star in the ASAS-3 survey ($P_{ASAS}=0.39725\,\mathrm{d}$)
coincides with our best-fitting period (see Table \ref{t:rrpars}). The ASAS V-filter observations are systematically brighter than ours, but this is due to
a bright neighbor ($V\sim13.6$) of star RRLS 26896, lying at a distance of $\sim14''$, which is
unresolved in the ASAS survey.
This is the only RRLS from the GCVS or the ASAS-3 that lies
in the surveyed region, and it was successfully recovered by our survey.

\section{EXPECTED NUMBER AND DISTRIBUTION OF GALACTIC RRLS}\label{s:rrmw}

RRLS are found in the halo and thick disk stellar populations of the MW, and a certain number of them are expected in the region that we have surveyed.  Below we estimate the numbers of these stars and their distributions in space and radial velocity to see if there is an excess of RRLS that should be attributed to another source. 

\subsection{Expected Radial Profile of Galactic RRLS}

Since the volume of our survey is naturally expressed in heliocentric Galactic 
coordinates $(D_\odot,l,b)$, the number $N_{RR}$ of RRLS in a given volume is expressed as

\begin{equation}
N_{RR}=\int_{D^o_\odot}^{D^f_\odot}\int_{b_o}^{b_f}\int_{l_o}^{l_f} \rho(D_\odot,l,b) D^2_\odot\:dD_\odot\, \cos{b}\,db\,dl
\label{nrr}
\end{equation}

This equation requires that the density profile $\rho$ be expressed in terms of heliocentric coordinates.
Since the density profiles of the halo and thick disk are naturally written in galactocentric 
coordinates, the coordinate transformation 
complicates the integral in Eq.\ref{nrr}, making extremely difficult to obtain an analytical solution. Consequently,
we solved it numerically by using a Monte Carlo implementation of the Von Neumann 
Rejection Technique \citep{pre02}.

The RRLS density profile of the Galactic halo has been well
constrained by several authors \citep[e.g.][]{viv06,wet96}. 
In the case of the Galactic thick disk, 
the RRLS density profile has only been characterized in the direction
perpendicular to the plane of the disk \citep{lay95}, which is due to the difficulties in surveying the
crowded fields near the Galactic plane where the interstellar extinction is high. 
The density profile of the thick disk
has been studied by several authors using different tracers and techniques 
\citep[e.g.][]{bro08,c-l05,lar03}, and we have adopted these profiles as valid for the RRLS too. 
The assumed analytical expressions for the 
density profiles of the Galactic halo and thick disk are listed in Table \ref{t:densprof}.
In Table \ref{t:densprof}, we also give the density profile of a hypothetically warped thick disk with the same warp properties as the thin disk. This profile is used in order to explore the
possibility that the thick disk might be warped, as suggested by \citet{car08}.
Since the possible warp of the thick disk has not been explored in detail, the best assumption
we can make is that it is described
by the same parameters as the warp of the thin disk \citep[e.g.][]{l-c02}.
This is a plausible assumption since analytical models of warp formation \citep{spa88}
and N-body simulations, which model the formation of warped
thin and thick disks \citep{qui93} predict similar properties.

The profile parameters, namely the RRLS density normalization $\rho_\odot^{RR}$ (for type ab RRLS), 
the power-law index $n$ and flattening ratio $c/a$ of the halo,
as well as the scale length $h_R$ and scale height $h_z$ of the thick disk, 
determined by several authors using different tracers are presented in
Table \ref{t:profpars}. Since the only determination available to date of $\rho_\odot^{RR}$ 
for the thick disk is that of \citet{lay95}, derived from nearby RRLS,  we use 
this value ($\rho_\odot^{RR}=10 \pm 4$ RRLS$/{\rm kpc}^3$) 
in the calculation of $N_{RR}$ using the scale lengths and heights 
derived by different authors (Table \ref{t:profpars}). 

\begin{figure}
\plotone{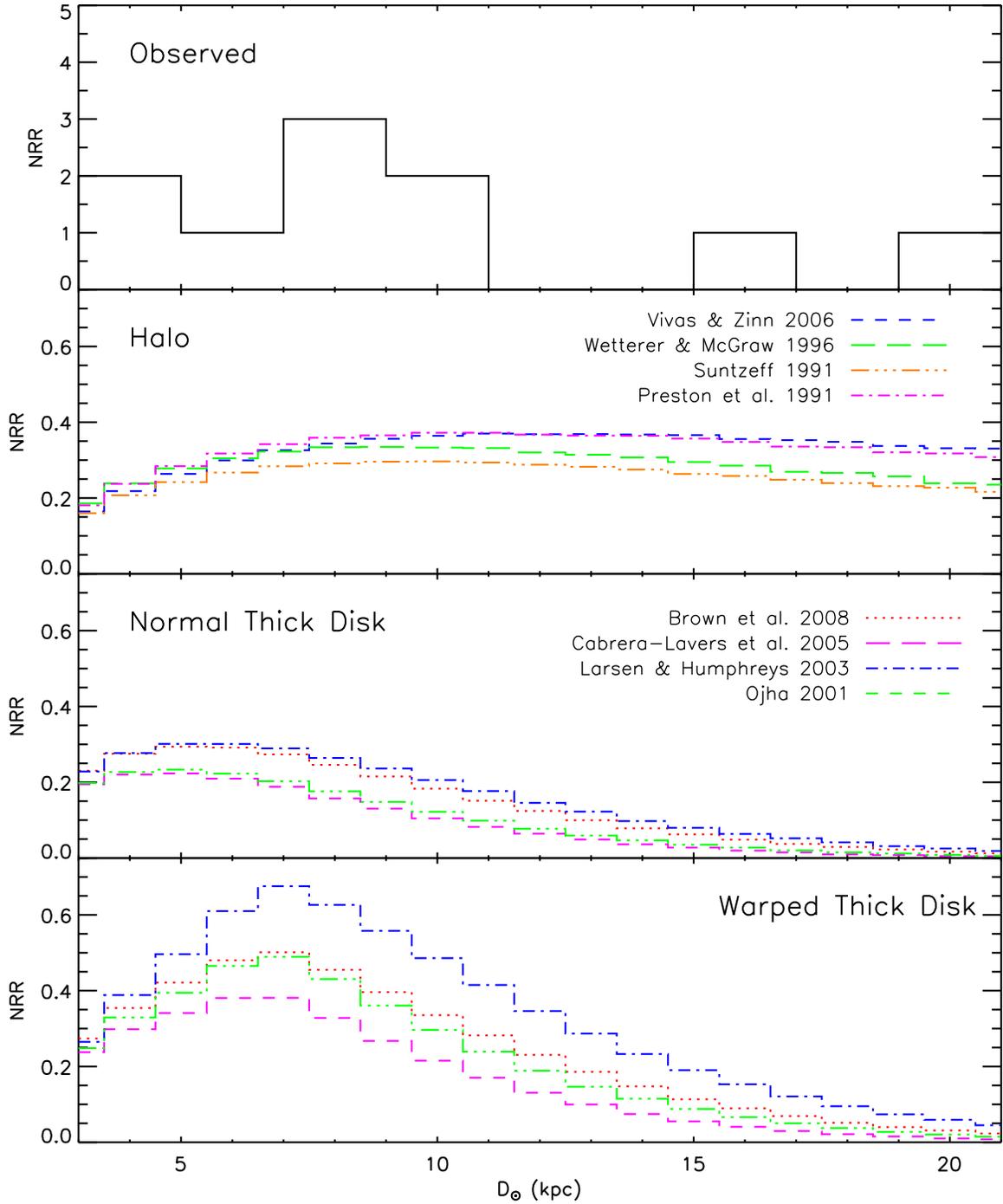}
\caption{Observed and expected numbers of RRLS as a function of heliocentric distance. From top to bottom: Observed distribution, Galactic halo, normal thick disk and warped thick disk distributions.}
\label{f:nrrprofile}
\end{figure}

By solving Eq. \ref{nrr} using the profiles given in Table \ref{t:densprof}, we obtained 
the total number of RRLS expected in the range $3\,\mathrm{kpc} \leq D_\odot \leq 20\, \mathrm{kpc}$. 
The results are shown in the last column of Table \ref{t:profpars}, in which $N_{RR}$
has been computed for RRLS of both types by correcting for the factor 
$N_{ab+c}/N_{ab}=1.29$ \citep{lay95} to account for RRLS of type c. 
Also, we derived the number of RRLS as a 
function of heliocentric distance, which is shown in Fig. \ref{f:nrrprofile} together with 
the observed radial distribution of RRLS in our survey.
From Fig. \ref{f:nrrprofile}, one can see that the theoretical RRLS radial distribution 
changes only slightly with the different sets of parameters. Most importantly, in the
case of the warped thick disk there is a peak located at $\sim 7.5$ kpc, independent
of the choice of the parameter set. This peak corresponds to the distance
where the survey's line of sight intersects the highest density part of the thick disk's warp.
The existence of this peak in the case of a warped disk is what led \citet{mom04,mom06}
and others \citep{l-c06,car08} to contest the interpretation of the CMa overdensity as a dwarf galaxy remnant, 
suggesting instead that it could be due to the warped Galactic disk(s). 

From Table \ref{t:profpars}, one can see that $4-6$ halo RRLS
are expected in the distance range of the survey, and $2-3$ or $3-6$ RRLS are expected
respectively from a normal or warped thick disk. This yields a total of $6-9$ or $7-12$ 
RRLS expected from a Galactic distribution. These numbers are in excellent agreement with
the $10$ RRLS found in our survey, leaving very little room (if any) for an excess of RRLS
over the Galactic background.

To take into account the random fluctuations that are obviously present in such small samples, 
we made $10^3$ random realizations of a halo plus a normal thick disk (H+TD) and of
a halo plus a warped thick disk (H+WTD). 
Each of these realizations was compared with the observed radial 
distribution, and the probabilities that the distributions 
are different were computed via a Kolmogorov-Smirnov test. These tests showed that 
for the H+TD and H+WTD distributions only $4\%$ and $1\%$, respectively, of the $10^3$ realizations 
were incompatible with the observed distribution (i.e. $P<0.01$). We conclude, therefore, 
that the observed RRLS radial distribution is consistent with the expectations from the Galactic models, 
although the small number of stars hinders any distinction between a normal and a warped thick disk.

Several studies have shown that there is a correlation between the [Fe/H] values of RRLS and their memberships 
in the stellar populations of the MW.  \citet{lay95}, for example, found that the majority of RRLS with [Fe/H]$<-1.3$ 
and [Fe/H]$>-1.0$ are members of the halo and thick disk populations, respectively, while the intermediate zone, 
$-1.3<$[Fe/H]$<-1.0$, is occupied by a mixture of these populations.  According to these limits, the probable 
memberships of the 8 RRLS in our sample that have [Fe/H] measurements are 4 halo stars, 1 thick disk star, and $3$ in
the mixture zone.  While these assignments must be viewed with caution because the [Fe/H] distributions 
of both populations have tails that extend beyond the region of large overlap, they are consistent with the 
results obtained from the density profiles.

\subsection{Comparison with the Observed RRLS Radial Velocity Distribution}\label{s:radvel}

\begin{figure}
\plotone{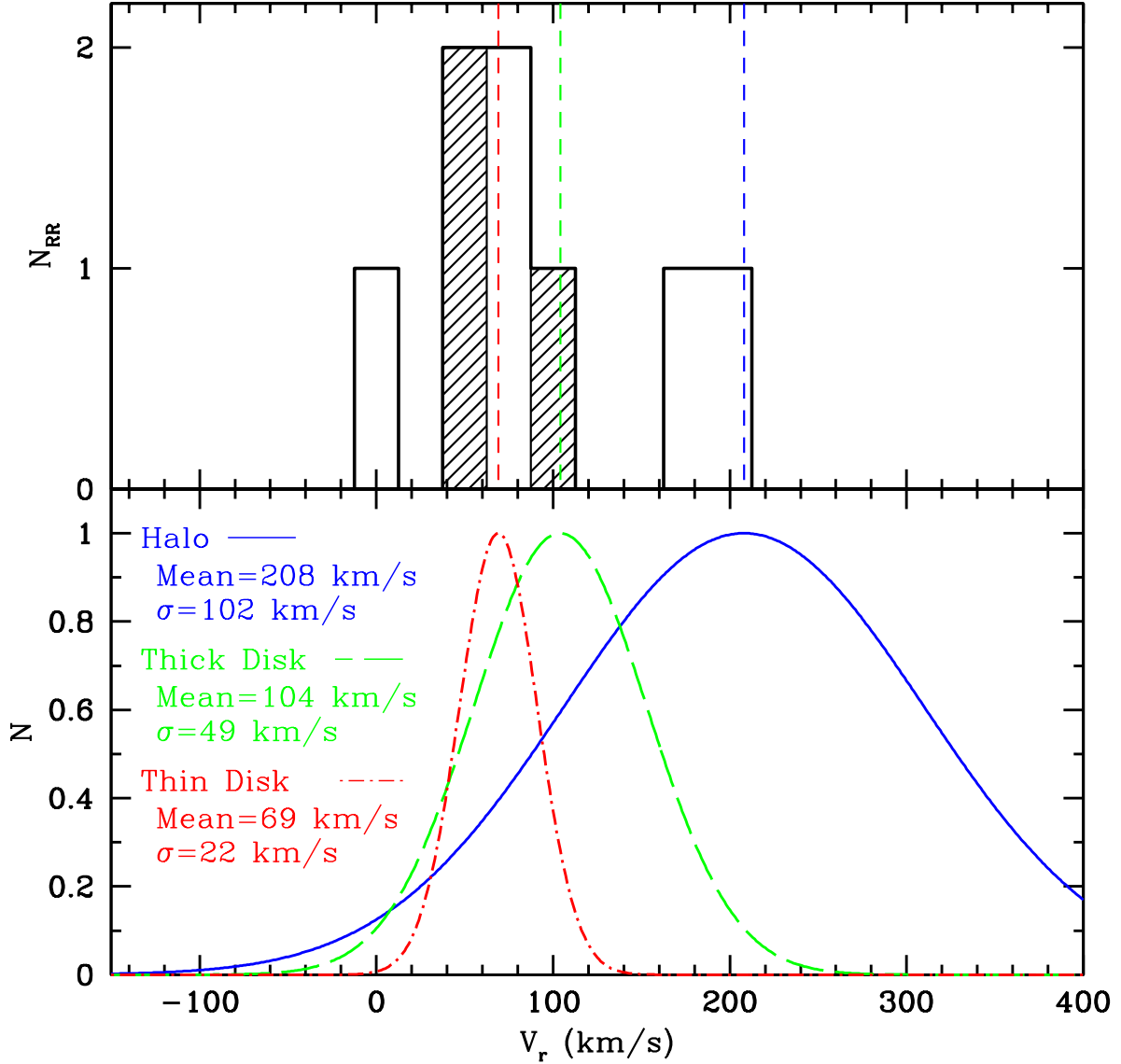}
\caption{\emph{Upper panel:} Radial velocity distribution of the 8 RRLS with $V<16.5$. The shaded histogram corresponds to the radial velocity distribution of RRc stars. \emph{Bottom panel:} Radial velocity distributions of thin disk A-F stars (dashed dotted line), thick disk (dashed line) and halo (solid line) stars in the surveyed area, taken from the Besan\c{c}on Galactic model \citep{rob03}. The dashed lines in the upper plot show the positions of the means of the distribution of each Galactic component.}
\label{f:radvel}
\end{figure}

As discussed above, we have measured the radial velocities of the 8 RRLS that lie in the distance range of the CMa overdensity ($ 4 > D_\odot > 12$
kpc). 
In the upper diagram of Fig.\ref{f:radvel}, the radial velocity distributions of the RRab and RRc variables are depicted by the clear and the shaded histograms, respectively.  They are plotted separately because the RRc sample is more prone to contamination from other types of stars.  In the lower diagram, we have plotted, with arbitrary normalizations,
the radial velocity distributions of A-F stars in the halo, thick, and thin disks, 
as given by the Besan\c{c}on Galactic model \citep{rob03} for the area of our survey. 
The halo and thick disk radial velocity distributions correspond to stars in the distance
range of the RRLS in our survey; while the thin disk distribution  
was obtained for stars without any constraint on distance.
The thin disk distribution is shown for reference since, as mentioned in \S \ref{s:contamination}, some contamination may remain and these non-RRLS are most likely members of the thin disk. The dashed lines in the upper plot show the positions of the means of the distribution
of each component.  From this figure one can see that the observed radial velocity distribution is consistent with the expected halo and thick disk distributions.  The radial velocities of some of the stars that we have classified as RRc are consistent with membership in the thin disk, but given the overlap in velocity with the halo and thick disk populations, it not certain that these stars are contaminants rather than true RRLS.  In conclusion, it appears that $100\%$ of the sample of RRLS can be explained by the contributions from the halo and thick populations and possible contaminants, and therefore, there is no excess of RRLS that requires another explanation. 

\citet{mar05} measured the radial velocities of large samples of red clump (RC) and red giant branch (RGB) stars in the CMa overdensity, and found that they define a sequence in a plot of $D_\odot$ against $V_r$ (see Figs. 6 \& 12  in \citet{mar05} and our Fig. 8).  Their sample of RGB stars were observed in several two-degree fields that were centered on $l = 240\degr$ and $-4\fdg 8 > b > -10\fdg 8$, and most of their sample of RC stars were observed in a field at (l, b) = ($240\fdg 0$, $-8\fdg 8$). Our RRLS survey encompasses the field where the majority of RC stars were observed, and it overlaps with many of the fields where the RGB stars were observed (see Fig. 1).  

\begin{figure}
\plotone{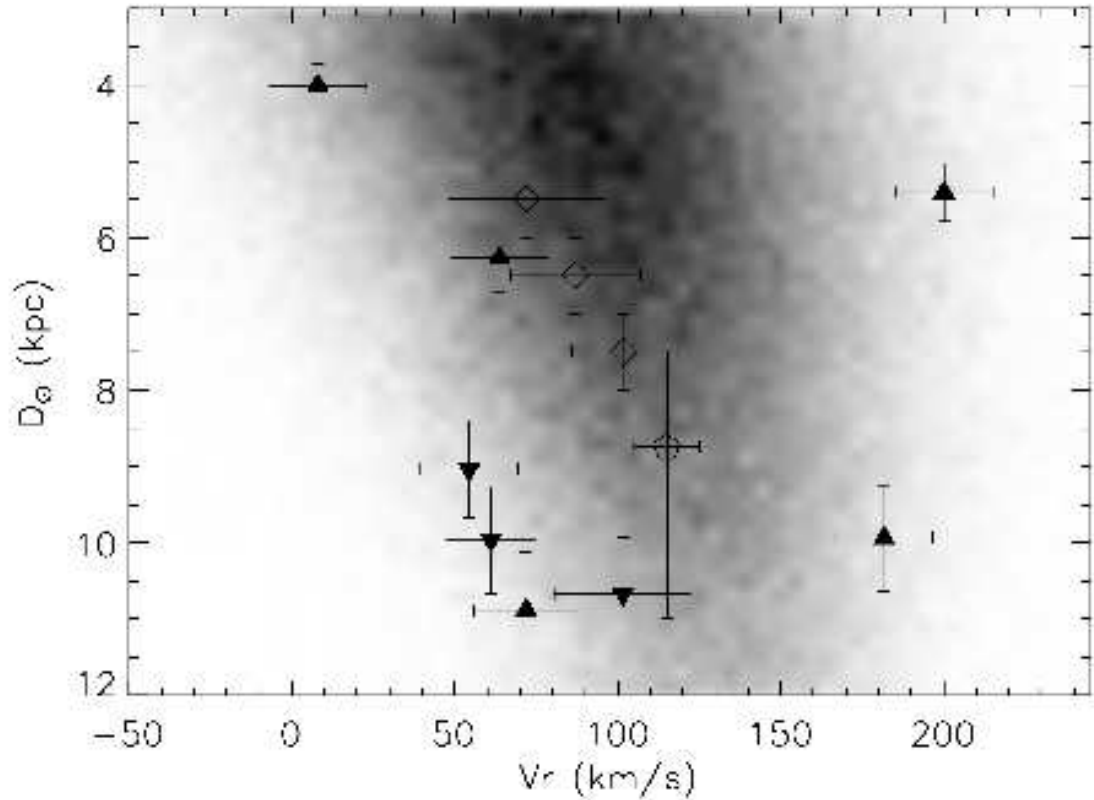}
\caption{Heliocentric distance versus radial velocity of the 8 RRLS with $V<16.5$. Solid triangles represent
RRLS of type ab (up) and c (down). Open symbols represent RC (rhombuses) and RGB (pentagon) stars identified with the CMa overdensity \citet{mar05}. The grayscale shows the distribution of thick disk stars in the surveyed area from the Besan\c{c}on Galactic model \citep{rob03}. The distribution of halo stars in this diagram (not shown) has a broad peak in Vr at $\sim200$ (see Fig. 8) and gentle falloff with $D_\odot$.}
\label{f:dvsvrad}
\end{figure}

In Figure~\ref{f:dvsvrad}, the sequence that is defined by the observations of \citet{mar05} is compared to our RRLS observations.  For the measurements of the RC stars, we adopted the mean values of $V_r$ that \citet{mar05} list in their Table 2 for different bins in $D_\odot$.  
In our plot of these values in Figure~\ref{f:dvsvrad}, the vertical error bars depict the size of the bin in $D_\odot$ (1 kpc for the RC stars), while the horizontal error bars depict the standard deviation of the Gaussians that \citeauthor{mar05} fit to the velocity peaks.  For the RGB stars, we adopted the mean $V_r$ and standard deviation that \citeauthor{mar05} give for the distance bin of $7.5 - 11$ kpc.   For the mean distance of these RGB stars, we adopted 8.5 kpc from Fig. 12 of \citet{mar05}.

Figure~\ref{f:dvsvrad} shows that several of the RRLS are widely scattered over the $V_r - D_\odot$ plane and that there is loose group at ($V_r$, $D_\odot$) $\sim$ (70 \kms, 10.5 kpc). This group should not given much weight because 3 of the 4 stars are type c, the type most likely contaminated by thin disk interlopers, which would have instead $D_\odot \lessapprox 4$ kpc.  Figure~\ref{f:dvsvrad} shows that the RRLS are not concentrated on the sequence defined by the observations of \citet{mar05}.  Two RRLS, the type ab variable 882 at ($V_r$, $D_\odot$) = (64 \kms, 6.3 kpc) and the type c variable 39928 at 102 \kms, 10.7 kpc, lie within the velocity and distance ranges of the RC and RGB stars that \citet{mar05} attribute to the CMa dwarf galaxy.  While this could mean that these RRLS belong to this system, this is not necessarily the case.  Members of the Galactic halo population are expected to have a very broad peak in velocity near 200 \kms and a roughly uniform distribution over the range of distances in Figure~\ref{f:dvsvrad} (see Figuress 6 \& 7).  Consequently, every RRLS in Figure~\ref{f:dvsvrad} is a candidate halo star.  The behavior of the Galactic thick disk population is depicted by the grayscale in Figure~\ref{f:dvsvrad}, which was obtained from the Besan\c{c}on model \citep{rob03}.  Since 882 and 39928 lie within the region expected to contain thick disk stars, it is possible that they belong to this Galactic population or the halo population and not to the CMa dwarf galaxy.   Our survey has detected 0 to 2 RRLS that \textit{could} be members of the CMa dSph galaxy

\section{THE NUMBER OF RRLS IN DWARF SPHEROIDAL GALAXIES}\label{s:dsphs}

Out of the 10 RRLS found by our survey, at most $2$ could be associated with the CMa dwarf galaxy.  This is many fewer RRLS than expected.

\begin{figure}
\plotone{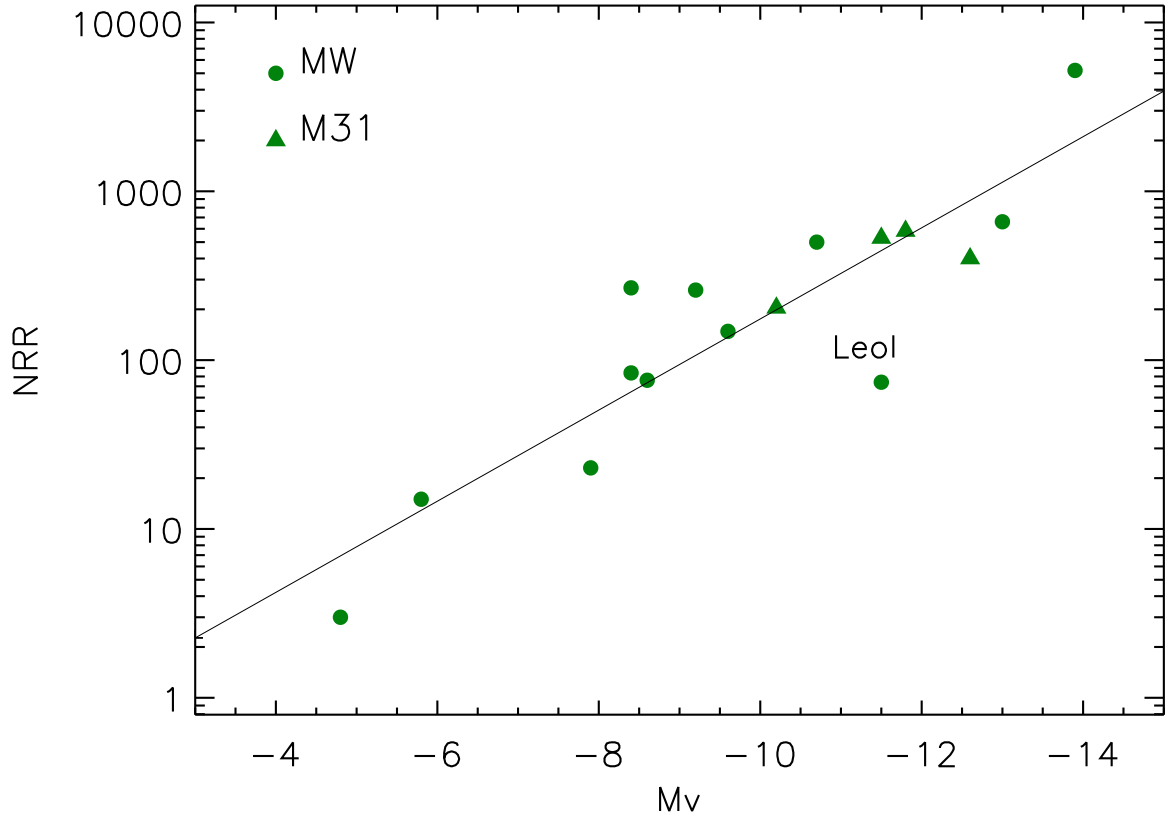}
\caption{The observed number of RRLS in dSph satellites of the MW (\emph{circles}) and M31 (\emph{triangles}).
A least squares fit to the data is represented by the straight line.  Note that $>1000$ RRLS are expected
in the CMa dSph galaxy if it has $M_V \sim -14$ \citep{mar04a,bel04,m-d05} and follows the trend in this diagram.
The position of Leo I, the MW dSph galaxy with the lowest $S_{RR}$ (see \S 4), is marked.}
\label{f:nrr_Mv}
\end{figure}

In Fig. \ref{f:nrr_Mv}, the number $N_{RR}$ of RRLS in
dSph satellite galaxies of the MW and M31 is plotted versus their absolute visual magnitude $M_V$ (data from
\citet{sie06}, \citet{kue08} and \citet{gre08}, and the compilation by \citet{viv06}).
There is a strong correlation between the
observed number of RRLS in a galaxy and its absolute magnitude in the sense that
the number of RRLS grows with increasing luminosity.
This correlation even holds for the extremely low luminosity satellites Boo, CVn I and II 
($M_V\gtrsim-8.5$). It also holds for the Sgr dwarf which is at present undergoing tidal disruption.
If the CMa dSph galaxy follows the same trend and has the estimated  
$M_V\sim-14.5$ \citep[estimates of the absolute V magnitude range 
from $-14.4$ to $-14.5$,][]{m-d05,bel06}, then it is expected to contain $\sim2880$ RRLS.  Our survey covers $\sim3\%$ of the galaxy, and therefore, it should have detected $\sim86$ RRLS. 
If CMa is more elongated with aspect ratio of $\sim6:1$, 
the coverage of our survey would be $\gtrsim2\%$ and hence $\gtrsim58$ RRLS should have
been detected by our survey. 
Since the expected number of RRLS is a factor of $\gtrsim40-30$ times greater than the number observed, 
we can safely reject the hypothesis that the CMa overdensity is caused by a "normal" dSph galaxy 
with $M_V\sim-14.5$.  We can also firmly reject the possibility that the luminosity of this galaxy has been simply overestimated.  Note that the trend in Fig. \ref{f:nrr_Mv} indicates that $\sim 600$ RRLS are expected in a "normal" dSph with $M_V\sim-12$, which is much fainter than any estimate for the CMa galaxy.  Nonetheless, if such a system were present, our survey should have detected $\sim 12$ RRLS as an excess over the expected Galactic background of RRLS, which is not observed.

It also clear from our survey that the putative CMa dSph galaxy must be unlike any other dSph galaxy in its production of RRLS per unit luminosity.
Fig. \ref{f:nrr_Mv} shows that the dSph galaxy that deviates by the largest amount below the line is the MW satellite Leo I. This was long considered to be an example of a galaxy whose stellar population is completely 
dominated by an intermediate-age population because there was no clear-cut evidence for an older population 
\citep[e.g.][]{lee93,gal99}.  The recent survey by \citet{hel01}, which discovered 74 RRLS 
in Leo I, showed that even this exceptional system has an old stellar population containing RRLS.  
The Leo I galaxy has the lowest specific frequency\footnote{The specif frecuency, $S_{RR}$, is defined
as the number of RRLS per unit absolute visual magnitude ($M_V$) normalized to $M_V=-7.5$ \citep{sun91}.} of RRLS ($S_{RR}$=1.9)
of any of the 13 dSph galaxies that are brighter than $M_V=-8$ \citep{viv06}.  If a dSph 
galaxy of $M_V=-14$ produces the CMa overdensity, the results of our survey suggest that it has $S_{RR} < 0.2$, an order of magnitude smaller than that of Leo I.

\section{DISCUSSION AND CONCLUSIONS}

The QUEST RR Lyrae survey in the CMa
overdensity has yielded the following results:

\begin{itemize}

\item Ten RRLS, $6$ of type $ab$ and $4$ of type $c$, were found 
by our survey, which spans $\sim20$ sq. deg. and is centered approximately at $(l,b)=(240\degr,-9.5\degr)$. 
The distance range covered by the RRLS is $3$ kpc $ \leq D_\odot \leq 20$ kpc, 
and $8$ out of the $10$ RRLS are within the distance range of CMa overdensity ($D_\odot<11$ kpc).

\item Integration of the density profiles of the Galactic halo and thick disk indicates 
that within the surveyed region the Galactic halo and thick disk should contribute a total of $6-9$ RRLS, and if the thick disk is warped like the thin disk, the expected number is $7-12$ RRLS. These numbers are in good agreement with the $10$ RRLS found by our survey; consequently there is no excess of RRLS in the CMa overdensity.

\item The radial velocities of 8 out of the 10 RRLS were measured, and their values are consistent with membership in either the Galactic halo or the thick disk populations.

\item An analysis of the number of RRLS found in the dSph satellites of the MW and M31 indicates that our survey should have discovered $\sim70$ RRLS if the CMa overdensity is produced by a typical dSph galaxy with $M_V=-14.5$, the estimated luminosity of the CMa system.  
Even if this galaxy is 2.5 magnitudes fainter, which is much lower than any estimate in the literature, the expected excess of RRLS is $\sim12$, which is not observed.  The CMa dSph galaxy must have a significantly lower value of the specific frequency of RRLS ($S_{RR}$) than Leo I, the Local Group dSph that has the lowest measured value.

\end{itemize}

The hypothesis that the CMa overdensity is produced by a dwarf galaxy is severely constrained by the above results.  
We have shown that if it is a dSph galaxy, the type favored by the proponents of the hypothesis, then it must be 
unlike any of the satellite dSph galaxies of the MW and M31.  No one has proposed that the putative galaxy 
is a dwarf irregular because there is no sign of a very young stellar population \citep[age $< 100$ Myr, e.g.][]{dej07}
or star forming regions associated with the CMa overdensity.  Nonetheless, we note that RRLS have been 
found in every dwarf irregular galaxy of the Local Group that has been adequately searched, and that there is 
evidence for very old and metal-poor stellar populations in the others \citep[see][and references therein]{cle03,mom05,mat98}.  These populations have probably produced RRLS that await detection.  

According to \citet{dej07}, the old stars in the CMD of the CMa overdensity have a mean metallicity 
of [Fe/H]$\sim -1$ and ages in the range 3-6 Gyrs, although they could not constrain the presence of an 
older population.  Clearly, the metals in this intermediate-age population must have come from earlier 
generations of more metal poor stars.  There is no sign of a blue horizontal branch (BHB) in the CMD, and our 
survey has shown that there are very few, if any, horizontal branch stars in the instability strip.  
The remaining alternative is that the majority of the oldest and most metal poor HB stars of the CMa galaxy 
must lie on the red side of the instability strip and are therefore part of the red clump that is observed in the CMD.
However, at low metallicities (i.e. [Fe/H]$<-2$) a relatively young age is required to keep the horizontal 
branch to red side of the instability strip \citep[see, for example, Fig. $9$ in][]{rey01}.  
Thus, the CMa dwarf must have started its star formation at a later time than is observed in other dwarf galaxies.  
This was once thought to be the case for Leo I, but we now know that this was incorrect \citep{hel01}.

\citet{mar04a} have suggested that the globular clusters NGC 1851, 1904, 2298, and 2808 once belonged to the 
CMa dwarf galaxy.  These clusters span a range in horizontal branch type from predominantly blue to predominantly red,
although each one contains many BHB stars and at least a few RRLS \citep{lee94,for04}.  The idea that they 
originated in the CMa dwarf is hard to reconcile with the lack of a BHB in the CMa CMD and our result that 
there is no excess of RRLS in the CMa overdensity.  In other dwarf galaxies that have globular cluster systems 
(e.g., the Sgr and Fornax dSph galaxies), there are stellar populations in the main body that resemble 
the ones in the retinues of globular clusters.    

In summary, the dwarf galaxy explanation for the CMa overdensity requires that the system be unique among dSph 
galaxies, and perhaps also dwarf irregulars, in its low production of RRLS.  The absence of RRLS in the 
overdensity is consistent with the alternative viewpoint 
that a combination of the thin and thick disk and spiral arm populations of the MW produces the CMa overdensity
\citep{mom04,mom06,l-c06,l-c07,car05,car08,moi06,pow08}.  
However, it does not prove that this alternative is correct, and more research on this question is warranted.

\acknowledgements

This research was based on observations collected at the J\"urgen Stock 1m Schmidt telescope and the 
1m Reflector telescope of the National Observatory of Llano del Hato Venezuela (NOV), which is operated by CIDA
for the Ministerio del Poder Popular para la Ciencia y Tecnolog{\'\i}a, Venezuela. 
The facilities of the 0.9m, 1.0m, 1.3m and 1.5m telescopes of the SMARTS Consortium at CTIO, 
Chile were also used.
C. Mateu acknowledges support from the predoctoral grant of the 
Academia Nacional de Ciencias F\'{\i}sicas, Matem\'aticas y Naturales of Venezuela.
C. Mateu would also like to thank Gustavo Bruzual and Gladis Magris for helpfull discussions
during the course of this research.  R. Zinn and L.R. Miller were supported by National Science Foundation Grant AST-05-07364.
The authors are gratefull for the assistance of the personnel, service-mode observers, 
telescope operators and technical staff at CIDA and CTIO, who made possible the acquisition of
photometric and spectroscopic observations at the NOV and SMARTS telescopes.
This work has also benefitted from use of the USNOFS Image and Catalogue Archive
operated by the United States Naval Observatory, Flagstaff Station
(http://www.nofs.navy.mil/data/fchpix/).

\clearpage

\clearpage

\clearpage

\clearpage

\clearpage

\begin{landscape}
\begin{deluxetable}{llllllll}
\tabletypesize{\scriptsize}
\tablecolumns{8}
\tablewidth{0pt}
\tablecaption{Summary of previous stellar population studies conducted in CMa.}
\tablerefs{B04=\citet{bel04}, M04=\citet{mar04a}, M05=\citet{mar05}, C05=\citet{car05}, Moi06=\citet{moi06}, B07=\citet{but07}, dJ07=\citet{dej07}, C08=\citet{car08}, P08=\citet{pow08}}
\tablehead{\colhead{Tracer} & \colhead{Age (Gyrs)} & \colhead{[Fe/H]} &\colhead{$D\odot$ (kpc)} & \colhead{Technique} & \colhead{Filters} & \colhead{Interpretation} & \colhead{Reference} }
\startdata
Blue Plume     & $1-2$            & \nodata              & \nodata            & MS-fitting         & BV    & dSph galaxy & B04 \\
Blue Plume     & $\sim0.7-2$      & $-0.3$ or $\sim-1.0$ & $\sim7.5$ or $9.3$ & HD modelling       & BR    & dSph galaxy & dJ07 \\
Blue Plume     & $\lesssim0.1$    & \nodata              & $10.8$             & TCDs, MS-fitting   & UBVRI & Spiral arm  & C05, Moi06 \\ 
Blue Plume     & \nodata          & $-0.37$ to $-0.5$    & $6.0 \pm 2.7$      & TCDs, spectroscopy & UBV   & Spiral arm  & P08 \\ 
B5-A0 stars    & $\lesssim0.1$    & \nodata              & $9.8^{+1.5}_{-1.0}$& TCDs               & UBVRI & Spiral arm  & C08 \\
\hline                                                               
MS             & $3-6$            & $\sim-1.0\pm0.1$     & $\sim7.5$          & HD modelling       & BR    & dSph galaxy & dJ07, B07 \\
RGB, MS        & $4-10$           & $-0.66$ to $-0.35$   & $\sim8.1 \pm 1.2$  & Isocrone fitting   & BV    & dSph galaxy & B04 \\
M-giants       & $4-10$           & $-1.0<[M/H]$         & $7.1 \pm 0.1$      & CMD fitting        & JHK   & dSph galaxy & M04 \\
RGB, RC        & \nodata          & \nodata              & $5.5$ to $8.5$     & Spectroscopy       &\nodata& dSph galaxy & M05 \\
F-G-K stars    & $6\pm2$          & $\sim-0.3\pm0.3$     & $6$                & TCDs               & UBVRI & Warped Thick Disk & C08 \\
\enddata
\label{t:cmastudies}
\end{deluxetable}
\end{landscape}

\begin{landscape}
\begin{deluxetable}{rlllllrllccccc}
\tabletypesize{\scriptsize}
\tablecolumns{10}
\tablewidth{0pt}
\tablecaption{Light curve and physical parameters of RRLS.}
\tablehead{
\colhead{ID} & \colhead{$\alpha$} & \colhead{$\delta$} & \colhead{$N$} & \colhead{Period} & \colhead{$Amp_{V}$} & \colhead{Type} & \colhead{$HJD_{max}$}  & \colhead{$\langle V_0\rangle$} &\colhead{$D_\odot$} &\colhead{$V_\gamma$ }&\colhead{$\sigma V_\gamma$ } & \colhead{[Fe/H]}\\
             & \colhead{(deg)}   & \colhead{(deg)}   &               & \colhead{$(d)$}  &                     &                & \colhead{$2450000\,d$}&                                   &\colhead{(kpc)}             &\colhead{(\kms)}&\colhead{(\kms)} & }
\startdata
7701& 108.429950& $-$28.128540 & 16& 0.58446& 0.71& RR$ab$& 3260.75930& 14.22 &   5.4 &  200   &  15    & $-$1.51\\
 882& 104.492882& $-$27.708120 & 29& 0.46353& 1.18& RR$ab$& 3372.89115& 14.54 &   6.3 &   64   &  15    & $-$1.52\\
20594& 105.830994& $-$27.565201 & 35& 0.50893& 1.32& RR$ab$& 3372.41209& 15.73 &  10.9 &   72   &  16    & $-$1.83\\
26896& 108.468697& $-$29.726669 & 29& 0.39725& 0.96& RR$ab$& 3372.85468& 13.57 &   4.0 &    8   &  15    & $-$1.29\\
29821& 108.591827& $-$31.073851 & 25& 0.37719& 0.25& RR$c$ & 3372.83819& 15.33 &   9.0 &   54   &  15    & $-$1.27\\
31038& 107.737091& $-$27.722931 & 25& 0.26672& 0.31& RR$c$ & 3372.95514& 16.73 &  17.2 & \nodata & \nodata& \nodata \\
32697& 107.872360& $-$27.761780 & 25& 0.20140& 0.46& RR$c$ & 3372.81509& 15.54 &  10.0 &   61   &  14    & $-$1.26\\
34757& 109.645363& $-$29.783131 & 20& 0.54532& 1.20& RR$ab$& 3372.59181& 17.09 &  20.3 & \nodata & \nodata& \nodata \\
39928& 109.702118& $-$30.578770 & 22& 0.33740& 0.33& RR$c$ & 3372.86688& 15.69 &  10.7 &  102   &  21    & $-$0.78\\
40676& 107.920067& $-$28.308350 & 31& 0.57517& 0.80& RR$ab$& 3373.13496& 15.54 &   9.9 &  182   &  15    & $-$1.44\\
\enddata
\label{t:rrpars}
\end{deluxetable}
\end{landscape}

\begin{landscape}\begin{deluxetable}{rlllllrll}\tabletypesize{\scriptsize}
\tablecolumns{10}
\tablewidth{0pt}
\tablecaption{Light curve parameters of rejected RRLS candidates.}
\tablehead{
\colhead{ID} & \colhead{$\alpha$} & \colhead{$\delta$} & \colhead{$N$} & \colhead{Period} & \colhead{$V_{amp}$} & \colhead{$HJD_{max}$}  & \colhead{$\langle V_0\rangle$} & \colhead{Reason for}\\
             & \colhead{(deg)}   & \colhead{(deg)}   &               & \colhead{$(d)$}  &                       & \colhead{$2450000\,d$}&                                &   \colhead{Rejection}}
\startdata
5564 & 105.416870& -28.277680 & 22& 0.29700& 0.17&  3258.68871& 15.69 & late-F spectra\\
10651& 105.439903& -31.074100 & 30& 0.37339& 0.37&  3372.46691& 13.88 &          W UMa\\
25711& 105.681969& -28.129801 & 32& 0.42189& 0.42&  3372.97294& 15.03 &          W UMa\\
6460 & 107.042080& -27.251020 & 22& 0.21343& 0.29&  3259.82428& 14.49 & late-F spectra\\
18056& 107.396880& -28.883540 & 19& 0.27842& 0.17&  3259.93843& 14.10 &          W UMa\\
17237& 108.275560& -26.897120 & 17& 0.24430& 0.24&  3259.75716& 14.36 & late-F spectra\\
\enddata\label{t:rejpars}
\end{deluxetable}
\end{landscape}

\begin{deluxetable}{lll}
\tablecolumns{2}
\tablewidth{0pt}
\tablecaption{Density profiles of the Galactic halo and thick disk.}
\tablecomments{In the table, $x,y,z$ are cartesian Galactocentric coordinates, $R_\odot=8$ kpc is the distance from the Sun to the Galactic center and $R_{gal}$ the Galactocentric distance along the plane. In the density profile assumed for a warped thick disk, $C_w=2.1\times10^{-19}$pc, $\varepsilon_w=5.25$ and $\phi_w=-5\degr$, are the warp parameters according to \citet{l-c02}.}
\tablehead{
\colhead{ } & \colhead{$\rho(\vec{r})$}  }
\startdata
{\bf Halo}       & $\rho_\odot^{RR} \Big(\frac{1}{R_\odot}\sqrt{x^2 + y^2 + (\frac{z}{c/a})^2}\Big)^n$ \\    \\ \hline \\
{\bf Thick Disk} & $\rho_\odot^{RR} \exp(-\frac{R_{gal}-R_\odot}{h_R})\exp(-\frac{|z|}{h_z})$ \\
\\ \hline \\
{\bf Warped}     & $\rho_\odot^{RR} \exp(-\frac{R_{gal}-R_\odot}{h_R})\exp(-\frac{|z-z_w|}{h_z})$ \\
{\bf Thick Disk} & 
with ${\scriptstyle z_w=[C_wR_{gal}(\mathrm{pc})^{\varepsilon_w}\sin(\phi-\phi_w)-15]\mathrm{pc}}$ if ${\scriptstyle R_{gal}<13\mathrm{kpc}}$\\
& with ${\scriptstyle z_w=z_w(13\mathrm{kpc})}$ if ${\scriptstyle R_{gal}>13\mathrm{kpc}}$\\
\enddata
\label{t:densprof}
\end{deluxetable}

\begin{landscape}
\begin{deluxetable}{llcccccc}
\tablecolumns{4}
\tablewidth{0pt}
\tabletypesize{\scriptsize}
\tablecaption{Number of RRLS expected in each Galactic component.}
\tablecomments{The symbol $N_{RR}$ represents the number of RRLS in the distance range from $3$ to $20$ kpc, 
both for the halo and normal/warped thick disk. {\footnotesize $^\dagger$For the scale-height Larsen \& Humphreys (2003) 
obtained $hz=870-930\pm50-80$pc, for the present 
study we use the intermediate value reported here.}}
\tablehead{
& \colhead{Reference}  & \colhead{$\rho_\odot^{RR}$ ($\frac{\#}{kpc^3}$)} & \colhead{$n$} & \colhead{$(c/a)$} &
\colhead{Tracer} & \multicolumn{2}{c}{$N_{RR}$}}
\startdata
{\bf Halo}  & \citet{viv06}    & $4.2^{+0.5}_{-0.4}$& $-3.1\pm0.1$    & Variable & RRLS & \multicolumn{2}{c}{$6$}  \\
     &\citet{wet96}           & $5.3\pm2$      & $-3.53 \pm 0.08$       & Variable & RRLS & \multicolumn{2}{c}{$4$}  \\
     &\citet{pre91}           & $4.8$              & $-3.2\pm0.1$    & Variable & RRLS & \multicolumn{2}{c}{$6$}  \\
     &\citet{sun91}           & $4.5\pm1$          & $-3.5$          & 1        & RRLS & \multicolumn{2}{c}{$5$}  \\
\hline
 & \colhead{Reference}  & \colhead{$\rho_\odot^{RR}$ ($\frac{\#}{kpc^3}$)} & \colhead{$h_z$ (pc)} & \colhead{$h_R$ (kpc)} & \colhead{Tracer} & \colhead{$N_{RR}^{normal}$} & \colhead{$N_{RR}^{warped}$}\\[6pt]
\hline
{\bf Thick} &\citet{bro08} & $\cdots$ & $1026 \pm 100$     & $3.5 \pm 0.5$       & BHB stars        &$3$ & $5$\\
{\bf Disk}  &\citet{c-l05} & $\cdots$ & $1061 \pm 52$      & $3.04\pm 0.11$      & 2MASS RC stars   &$2$ & $3$\\
           &\citet{lar03}$^\dagger$ & $\cdots$ & $900 \pm 65$       & $4.7 \pm 0.2$       & Star counts      &$3$ & $6$\\
           &\citet{ojh01} & $\cdots$ & $860 \pm 200$      & $3.7^{+0.8}_{-0.5}$ & 2MASS star counts&$2$ & $4$\\
\enddata
\label{t:profpars}
\end{deluxetable}
\end{landscape}

\end{document}